\documentclass[reprint,aps,twocolumn]{revtex4-1}%revtex4-1
\usepackage[unicode]{hyperref}
\usepackage{amsmath}
\usepackage[usenames]{color}
\usepackage{color}
\usepackage{colortbl}
\definecolor{linkcolor}{rgb}{0.8,0,0.2}
\definecolor{citecolor}{rgb}{0,0.6,0.2}
\definecolor{urlcolor}{rgb}{0,0,1}
\hypersetup{
    colorlinks, linkcolor={linkcolor},
    citecolor={citecolor}, urlcolor={urlcolor}}
\usepackage{graphicx}
\usepackage{epstopdf}
\usepackage[utf8]{inputenc}
\usepackage[english]{babel}

\usepackage{amsfonts}
\usepackage{eufrak}
\usepackage{mathrsfs}
\usepackage{bm}
\begin{document}
\title{Tunable Tamm plasmon polariton based planar hot-electron photodetector \\ from
$O-$ to $U-$ telecommunication band}
\author{Yurii V. Konov$^{1,2}$}
\author{Dmitrii A. Pykhtin$^{1}$}
\author{Rashid G. Bikbaev$^{1,2}$}
\author{Ivan V. Timofeev$^{1,2}$}
\affiliation{$^1$Kirensky Institute of Physics, Federal Research Center KSC SB
RAS, 660036, Krasnoyarsk, Russia}
\affiliation{$^2$Siberian Federal University, Krasnoyarsk 660041, Russia}

\date{\today}

\begin{abstract}
Developing tunable photodetectors that can operate over a wide range of wavelengths and integrating them into integrated circuits is a significant challenge in today's technology. These devices must be miniaturized, inexpensive, and easily manufactured. To address this challenge, we propose the development of a tunable planar hot-electron photodetector based on Tamm plasmon polariton.
The tuning of the operational wavelength is achieved by incorporation of a material with a phase transition, Sb$_2$S$_3$, into Tamm plasmon polariton based structure. This allows for tuning of the detection wavelength over a broad range, encompassing all bands of the telecommunications spectrum.

\end{abstract}

\maketitle

\section{Introduction}

In recent years, there has been a growing interest in Tamm plasmon-polaritons (TPPs), which are light states localized at the interface between a one-dimensional photonic crystal (PhC) and a metal film~\cite{Kaliteevski2007,Kar2023,Chen_2023_TPP_GS_APR,Chen_2022_TPP_UL_APR,Bikbaev_2020_HMM_JOSAB,nano13040693,Bikbaev2022,Wu2023,Wu20231,Konov2024}. The wavelength and Q-factor of TPPs can be controlled by introducing a liquid crystal into the metal-photonic crystal structure \cite{Cheng2018}, which allows for control over the optical properties of the system by simultaneously varying the electric field and temperature.
However, devices based on this approach are relatively slow, since the response time of liquid crystals is at least one millisecond. A promising alternative is a phase change materials such as VO$_2$ \cite{Liu2023,Wen2024,Fan2022}, GeSbTe (GST) \cite{Raoux2009,Cao2020,9133093}, and Sb$_2$S$_3$ \cite{Delaney2020,Farhana2023,Gao2024}. These materials exhibit a sharp change in their optical properties at a specific temperature, allowing for rapid modulation of the system's optical response. In this case, the switching occurs in a microsecond, three orders of magnitude faster than in liquid crystal-based structures.
The advantage of VO$_2$ is the low phase transition temperature at 68~C$^o$.
However, like GST, VO$_2$ has a high extinction coefficient, which makes it difficult to use in nanophotonic devices.

The advantage of Sb$_2$S$_3$, compared to VO$_2$ and GST, is the significant change in its refractive index ($\Delta n \approx 1$ at a wavelength of 600 nm) due to phase transition from an amorphous to a crystalline state. Moreover, the extinction coefficient at the wavelength above 700~nm is close to zero, which makes this material promising for use in telecommunication devices~\cite{Ye2025}.
It is important to note that this material also demonstrates excellent switching ability~\cite{Lawson2023}. This material can undergo up to 1,000,000 cycles of switching from amorphous to crystalline state and back, without any deterioration in its optical properties. This makes them promising candidates for designing controlled nanophotonic devices, such as solar cell~\cite{Kondrotas2018} or wavelength-tunable photodetector~\cite{Santos2022}. In~\cite{Santos2022} the authors proposed a model for a tunable photodetector based on a two-dimensional lattice of gold nanostrips deposited on the Sb$_2$S$_3$ substrate. They showed that switching Sb$_2$S$_3$ between the amorphous and crystalline phases leads to a shift in the resonant wavelength from $O-$ to $C-$ band. However, the fabrication of such 2D gratings is a challenging and expensive process.
Therefore, in this paper, we propose a simple and easy-to-manufacture model of a tunable photodetector for the telecommunication range based on TPP. We show the possibility of tuning a TPP wavelength by switching Sb$_2$S$_3$ and by varying the angle and polarization of light incident on the structure.

\begin{figure*}
    \centering
    \includegraphics[scale=4]{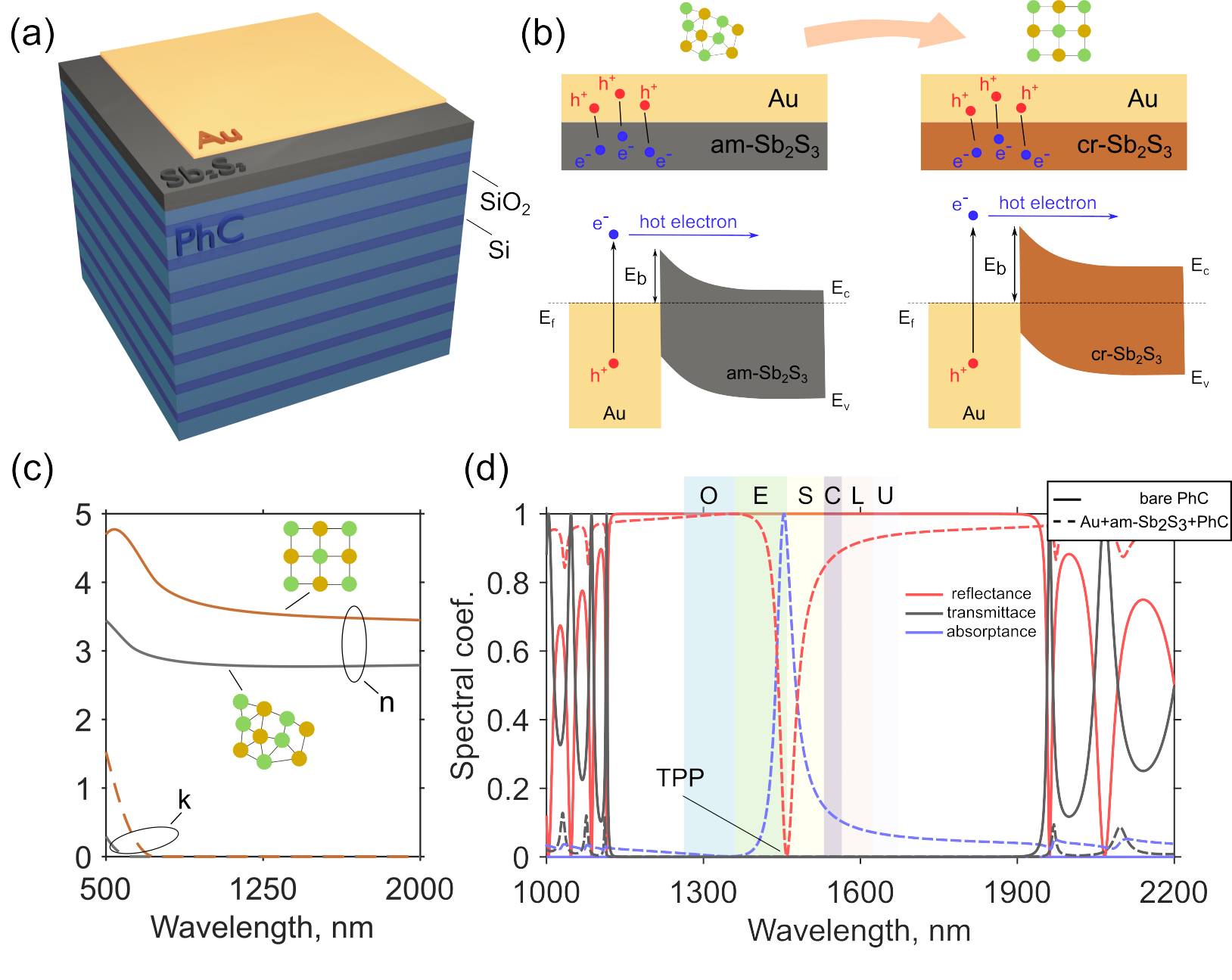}
    \caption{(a,b) Schematic representations of the proposed structure and Schottky barrier at the metal-semiconductor interface. (c) Wavelength dependencies of the real and imaginary parts of the complex refractive index of Sb$_2$S$_3$ in the amorphous (am-Sb$_2$S$_3$) and crystalline (cr-Sb$_2$S$_3$) phases. (d) Reflectance spectra of the bare PhC and the PhC conjugated to the Sb$_2$S$_3$ layer and the metal film.}
    \label{fig:fig1}
    \end{figure*}

\section{Description of the model}

\begin{figure*}
    \centering
    \includegraphics[scale=3]{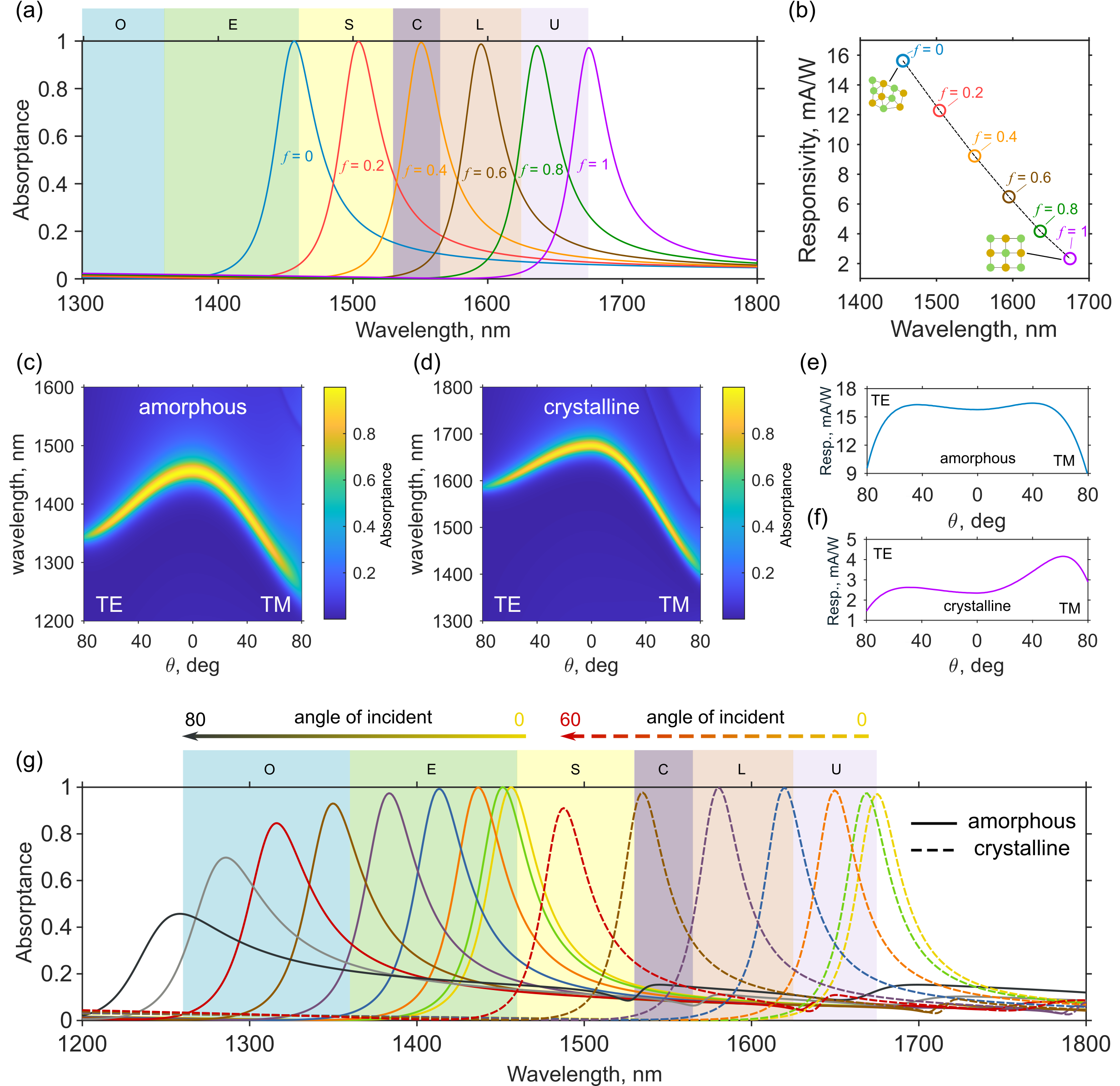}
    \caption{(a) Absorptance spectra of the structure at different values of $f$. (b) The dependence of the device responsivity from wavelength and $f$. Absorptance spectra of the structure for (c) amorphous and (d) crystalline phase of Sb$_2$S$_3$. Dependence of the responsivity of the device on the angle of incidence of TE- and TM-waves for (e) amorphous and (f) crystalline phase of Sb$_2$S$_3$. (g) Absorptance spectra of the structure for amorphous and crystalline phase of Sb$_2$S$_3$ at different values of the angle of the TM-wave incidence on the structure.}
    \label{fig:fig2}
    \end{figure*}
    
A schematic representation of the structure is shown in Figure~\ref{fig:fig1}. The PhC consists of silicon dioxide (SiO$_2$)~\cite{Malitson1965} and silicon (Si)~\cite{Palik1985} layers with thicknesses of 250 nm and 100 nm, respectively. The number of PhC layers $N=$ 12. 
A 150 nm thick layer of $Sb_2S_3$~\cite{Ye2025} were deposited on the surface of the PhC, followed by a gold layer~\cite{Johnson1972}.
The dependencies of the real and imaginary parts of the complex refractive index of Sb$_2$S$_3$ in the amorphous and crystalline phases are presented in Figure~\ref{fig:fig1}. It can be seen that at wavelengths longer than 700 nm the imaginary parts of the refractive indices are equal to 0. At the same time, the real part of the refractive index increases during the transition from amorphous to crystalline phase. In telecommunication region ($\lambda=$ 1550 nm) the real part of the refractive index increases from 2.77 to 3.49. 

Reflectance spectra of the bare PhC calculated by the transfer-matrix method~\cite{Yeh1979}, presented in Figure~\ref{fig:fig1}(d) by a solid line. The thicknesses of the PhC layers were chosen so that its band gap overlaps all bands of telecommunication spectrum (from $O-$ to $U-$band). 
Conjugated of PhC with am-Sb$_2$S$_3$ and gold films leads to excitation at their interface of TPP. As a result, a narrow spectral line appears in the reflectance spectra of the structure.
The near-zero reflection coefficient at the resonant wavelength and, as a consequence, the almost 100\% absorption is due to the fulfillment of the critical coupling condition of the incident field with the TPP. 
The structural parameters that satisfy this condition can be predicted using the temporal coupled mode theory~\cite{Haus1985}, as described in our previous work~\cite{Konov2024}. In the proposed design, energy from a localized state leaks out through three channels: transmission through the PhC, absorption by the metal film, and transmission through the metal.
In our case, the PhC is non-transparent, and all energy loss occurs via the transmission and absorption of the gold film. When the energy loss rates via these two mechanisms are equal, the critical coupling condition has been reached.
For the presented structure, this condition is satisfied for a gold film with a thickness of 20 nanometers. 

The tuning of the TPP wavelength in the PhC band gap was carried out using the phase matching condition~\cite{Kaliteevski2007,Vyunishev2019BroadbandPolariton}, which includes three parameters: $\varphi_{Au-Sb_2S_3}$, $\varphi_{top}$, and $\varphi_{PhC}$.
The phase of reflection from the metal-Sb$_2$S$_3$ pair depends on its thickness, while the phase of reflection from PhC depends on the number of layers. Changing these parameters would lead to a violation of the critical coupling condition, so the phase matching condition can only be achieved by varying the thickness of the first layer of PhC. Calculations have shown that for TPP excitation at the $E-$ and $S-$band boundaries, the thickness of the first layer of PhC should be 110~nm (see Fig.~\ref{fig:fig1}(d), dashed line).
Switching Sb$_2$S$_3$ from an amorphous to a crystalline phase leads to an increase in the phase shift in this layer, and as a result, a TPP wavelength shifts to long-wave region of the spectrum. The absorptance spectra of the structure for various values of $f$ (fractions of crystalline phase) are shown in Figure ~\ref{fig:fig2}(a). The refractive index of Sb$_2$S$_3$ was calculated using an effective medium theory in the range of $f$ from 0 to 1:

\begin{gather}
b=(3f-1)\varepsilon_{cr}+(2-3f)\varepsilon_{am}, \\ 
%n_{Sb_2S_3}= \sqrt{(b+\sqrt{8\varepsilon_{am}\varepsilon_{cr}+b^2})/4}.  \\
n_{Sb_2S_3}= \sqrt{\frac{b+\sqrt{8\varepsilon_{am}\varepsilon_{cr}+b^2}}{4}}.
\end{gather}

The Figure shows that the TPP wavelength shifts from 1450~nm at $f =$ 0 to 1675~nm at $f =$ 1 completely crossing $S-$, $C-$, $L-$ and $U-$band. It should be noted that as $f$ increases, a slight decrease in the absorption coefficient at the resonant wavelength is observed. This effect is explained by the violation of the critical coupling condition caused by the gold film reflection coefficient increase with wavelength growth. 

The proposed structure can be used as a tunable hot electron photodetector. 
The light absorbed by the gold layer excites hot electrons, which then move to the metal-semiconductor interface. When electron energy exceeds the Schottky barrier the electron can be injected into the semiconductor film and contribute to the photocurrent. Therefore, the photocurrent directly depends on the absorption coefficient at the resonant wavelength and also on the free path length of the electrons, which limits the thickness of the metal layer. In a metal layer thicker than 20~nm, the injection of hot electrons is reduced due to scattering within the conductor. At a gold film thickness of 20~nm in the proposed structure, critical coupling is achieved, allowing for neglecting electron scattering as they move towards the metal-semiconductor interface, resulting in a high probability of hot electron injection into the semiconductor. In this case, the responsivity of the device can be calculated using the Fowler model:

\begin{equation}
       R = \frac{qA\eta}{h\nu}, \quad \eta = \frac{(h\nu-\Delta E_b)^2}{4E_f h\nu},
\label{eq.eq1}
\end{equation}
where $q$ is the elementary charge, $A$ is the absorption coefficient at the resonant wavelength, $\eta$ is the quantum efficiency, and $h\nu$ is the photon energy. 
The Fermi energy of gold E$_f = 5.5$~eV. The Schottky barrier height E$_b$ at the boundary between gold and amorphous Sb$_2$S$_3$ is 0.35~eV, and at the boundary between gold and crystalline Sb$_2$S$_3$ is 0.57~eV~\cite{Santos2022} . 
The responsivity of the proposed device calculated at the resonant wavelength at different values of $f$ is shown in Fig.~\ref{fig:fig2}(b).
It can be seen that as the fraction of the crystalline phase increases, the responsivity decreases from 15.6 mA/W to 2.3 mA/W.

The strong dependence of the TPP wavelength on angle and polarization of the incident light can be used for the creation of a device with angular sensitivity. Figures~\ref{fig:fig2} (c-d) show the absorptance spectra of structures based on amorphous and crystalline Sb$_2$S$_3$ for TM- and TE-polarized waves.
As the angle increases, the resonant lines shift towards the short-wavelength region of the spectrum. For TM-waves, the blue shift of the resonant line is greater than for TE-waves. Additionally, the absorptance at the resonant wavelength remains close to unity over a wide range of angles for both polarizations, and decreases at angles greater than 50 degrees.
Figure \ref{fig:fig2}(g) shows the absorptance spectra of the structure in amorphous and crystalline phases of Sb$_2$S$_3$ for different angles of incidence TM-wave. Switching the Sb$_2$S$_3$ with varying the angle of incidence, allows to tune the TPP wavelength between 1260 and 1675 nanometers, completely covering all bands of the telecommunication spectrum.

According to the Fowler model, the magnitude of the photocurrent is directly proportional to the absorptance coefficient. This means that as the angle of incidence gets larger, the device's responsivity remains close to its value at normal incidence. Only when the angle exceeds 50 degrees does it start to decrease.
To verify this hypothesis, we calculated the angular dependence of the responsivity for both the amorphous and crystalline phases of Sb$_2$S$_3$. The results are shown in Figures~\ref{fig:fig2} (e-f).
For am-Sb$_2$S$_3$ , the responsivity increases with increasing angle for both polarization states. For the TM wave, it reaches its maximum at 40 degrees (16.4 mA/W), and for the TE wave at 44 degrees (16.3 mA/W).
For cr-Sb$_2$S$_3$  phase, the responsivity also increases with angle. For the TM wave, it reaches its maximum at 62 degrees (4.15 mA/W), and for the TE-wave at 50 degrees (2.62 mA/W).
This effect can be explained by the fact that, as the angle increases above 40 degrees, the absorption coefficient approaches unity. At this point, the TPP wavelength undergoes a blue shift. This means that increasing the frequency of a localized state, with a relatively small change in absorption coefficient, results in increased responsivity. However, at angles greater than about 60 degrees, increasing the frequency does not offset the sharp decrease in absorption coefficient, and responsivity rapidly decreases.

\section{Conclusions}

A model of a tunable planar hot electron photodetector based on the Tamm plasmon-polariton for the telecommunication region of the spectrum is proposed. The tuning of the operating wavelength of the device was realized by introducing on the metal-photonic crystal boundary Sb$_2$S$_3$ - a material with a phase transition from amorphous to crystalline state. It is shown that at normal incidence of radiation on the structure, the switching of Sb$_2$S$_3$ leads to a shift of the wavelength of the Tamm plasmon-polariton into the long-wavelength region of the spectrum, completely crossing the $S$,$C$,$L$ and $U-$ bands. The responsivity at 1550~nm reaches 9.22~mA/W. The absorption spectra of the structure at oblique incidence of TE- and TM-type waves are investigated. It is shown that for TE-type waves the change of the angle of incidence with the switching of Sb$_2$S$_3$ from amorphous to crystalline phase, makes it possible to tuning the resonant wavelength from the $O-$ to $U-$ band of the telecommunication region of the spectrum. The obtained results can be useful in designing nanophotonic devices based on the Tamm plasmon-polariton with improved spectral characteristics.

\bibliography{main.bbl}

\end{document}